\documentclass[10pt]{article}

\usepackage[top=1in, bottom=1.25in, left=1.25in, right=1.25in]{geometry}

\usepackage{setspace}
\usepackage{algorithm}

\usepackage{algpseudocode}

\usepackage{amssymb}

\usepackage{amsthm}

\usepackage{pifont}

\usepackage{amsmath}

\usepackage{adjustbox}

\usepackage{bm}

\usepackage{bbm}

\usepackage{natbib}

\usepackage{graphicx}

\usepackage{booktabs}

\usepackage{blkarray}

\usepackage{titlesec}  

\usepackage{authblk}

\usepackage[labelfont=bf]{caption}

\titleformat*{\section}{\sffamily\large}
\titleformat*{\subsection}{\sffamily\normalsize}

\newcommand{\mat}[1]{\bm{#1}}
\newcommand{\vect}[1]{\bm{#1}}
\newcommand{\set}[1]{\left\lbrace #1 \right\rbrace}
\newcommand{\T}{\mathsf{T}}

\newtheorem{definition}{Definition}

\newtheorem*{theorem*}{Theorem}

\title{On missing label patterns in semi-supervised learning}

\author[ ]{Daniel Ahfock}
\author[ ]{Geoffrey J. McLachlan}
\affil[ ]{School of Mathematics and Physics, University of Queensland, Brisbane, Australia}
\date{}                     
\begin{document}
\maketitle

\begin{abstract}
We investigate model based classification with partially labelled training data. In many biostatistical applications, labels are manually assigned by experts, who may leave some observations unlabelled due to class uncertainty. We analyse semi-supervised learning as a missing data problem and identify situations where the missing label pattern is non-ignorable for the purposes of maximum likelihood estimation. In particular, we find that a relationship between classification difficulty and the missing label pattern implies a non-ignorable missingness mechanism.  We examine a number of real datasets and conclude the pattern of missing labels is related to the difficulty of classification. We propose a joint modelling strategy involving the observed data and the missing label mechanism to account for the systematic missing labels. Full likelihood inference including the missing label mechanism can improve the efficiency of parameter estimation, and  increase classification accuracy.
\end{abstract}

\section{Introduction}
Semi-supervised learning tasks involve the analysis of datasets that are comprised of both labelled and unlabelled training observations.  Semi-supervised learning can be studied using the missing data framework of \citet{RUBIN1976}, where the unobserved labels are treated as the missing data. Previous theoretical work on semi-supervised learning involves the critical assumption that the missing data process is ignorable for the purposes of likelihood based inference \citep{mclachlan_1975_iterative, mclachlan_1977_estimating, ONeill1978,  zhang_2000_value, Chawla2005}. In many practical semi-supervised learning tasks, labels are manually assigned by domain experts. In these situations, a missing label can indicate that a particular observation is difficult to classify. We argue that this phenomenon strongly implies a non-ignorable missingness mechanism. A consequence is that fully efficient maximum likelihood inference will then require a model for the missing labels. 

To elaborate, suppose the dataset consists of $n$ observations that can be divided into $g$ classes, and we are interested in performing model based clustering or classification. At the population level, we have random features $\vect{X}_{i}$ and class assignments $\vect{Z}_{i}$, for $i=1, \ldots, n$.  Each cluster indicator vector $\vect{Z}_{i}$ is a $g$-dimensional random vector  $\vect{Z}_{i}=(Z_{i1}, \ldots, Z_{ig})^{\T}$. Element $h$ in $\vect{Z}_{i}$ is equal to one if observation $i$ is in group $h$ and is zero otherwise, $(i=1, \ldots, n; h=1, \ldots, g)$. The underlying data generating process is taken to be a finite mixture model. Let $\vect{\pi}=(\pi_{1}, \ldots, \pi_{g})$ give the mixing weights, where $\sum_{h=1}^{g}\pi_{h}=1$. Let $\vect{\theta}_{h}$ give the parameter for the $h$th component density, $f(\cdot \ ; \vect{\theta}_{h})$, for $h=1, \ldots, g$. The complete set of parameters for the mixture model is given by $\vect{\Psi}=(\vect{\pi}, \vect{\theta}_{1}, \ldots, \vect{\theta}_{g})$. Independently for $i=1, 
\ldots, n$, we have the hierarchical model:
\begin{align*}
    \vect{Z}_{i} &\sim \text{Multinomial}(1, \vect{\pi}), \\
    \vect{X}_{i} \mid \vect{Z}_{ih}=1 &\sim f(\vect{x}_{i}; \vect{\theta}_{h}).
\end{align*}
The marginal distribution of the features is given by $f(\vect{x}_{i} ; \vect{\Psi}) = \sum_{h=1}^{g}\pi_{h}f(\vect{x}_{i}; \vect{\theta}_{h})$. We assume the existence of some missing data mechanism so that not all labels are observed in the sample dataset.

For semi-supervised learning, we have $n_{1}$ labelled observations and $n_{2}$ unlabelled observations.  Let $\vect{x}_{j}^{(1)}$ refer to the observed feature vector for the $j$th labelled observation for $j=1, \ldots, n_{1}$. Let $\vect{x}_{k}^{(2)}$ refer to the observed feature vector for the $k$th unlabelled observation for $k=1, \ldots, n_{2}$. Let $\vect{x}^{(1)}=(\vect{x}_{1}^{(1)},\ldots, \vect{x}_{n_{1}}^{(1)})$ denote the features for the $n_{1}$ labelled observations and let $\vect{z}^{(1)}=(\vect{z}_{1}^{(1)}, \ldots, \vect{z}_{n_{1}}^{(1)})$ refer to the observed labels for the $n_{1}$ labelled observations. Similarly, let $\vect{x}^{(2)}=(\vect{x}_{1}^{(2)}, \ldots, \vect{x}_{n_{2}}^{(2)})$ refer to the observed features for the $n_{2}$ unlabelled observations. Ignoring the missing label mechanism, we can form a likelihood function using the available dataset $(\vect{x}^{(1)}, \vect{z}^{(1)}, \vect{x}^{(2)})$:
\begin{align}
\mathcal{L}_{\text{ign}}(\vect{\Psi}; \vect{x}^{(1)}, \vect{z}^{(1)}, \vect{x}^{(2)}) &=\left( \prod_{j=1}^{n_{1}}\prod_{h=1}^{g}[\pi_{h}f(\vect{x}^{(1)}_{j} ; \vect{\theta}_{h})]^{z^{(1)}_{jh}}\right) \left( \prod_{k=1}^{n_{2}}\sum_{h=1}^{g}\pi_{h}f(\vect{x}^{(2)}_{k} ; \vect{\theta}_{h}) \right). \label{eq:ignorance}
\end{align}
The subscript is used to emphasise that the missing label mechanism is ignored. An important theoretical concern is the suitability and efficiency of \eqref{eq:ignorance} for likelihood based inference.

Our main conclusion is that if missing labels are associated with observations that experts found challenging to classify, the pattern of missingness can then be relevant for maximum likelihood estimation. In a model based framework, the difficulty of classifying an observation can be quantified using the Shannon entropy. Let $\tau_{ih}$ represent the posterior probability that observation $i$ belongs to group $h$, given feature $\vect{x}_{i}$:
\begin{align}
    \tau_{ih} &= \dfrac{\pi_{h}f(\vect{x}_{i} ; \vect{\theta}_{h})}{\sum_{r=1}^{g}\pi_{k}f(\vect{x}_{i} ; \vect{\theta}_{r})}, \quad h=1, \ldots, g; \ i=1, \ldots, n. \label{eq:posterior_class_intro}
\end{align}
The classification Shannon entropy, $e_{i}$, is a function of the posterior class probabilities:
\begin{align}
    e_{i} &= -\sum_{h=1}^{g}\tau_{ih} \log \tau_{ih}, \quad  i=1, \ldots, n. \label{eq:shannon_entropy_intro} 
\end{align}
The Shannon entropy takes values in the interval $[0, \log g]$, and observations near class boundaries will have greater entropy. We propose to model the missing label probability as a function of the Shannon entropy. As the classification entropy is a function of the model parameters $\vect{\Psi}=(\vect{\pi}, \vect{\theta}_{1}, \ldots, \vect{\theta}_{g})$, the missing data model provides supplementary information for the estimation of $\vect{\Psi}$.  Joint modelling can be carried out using a profile likelihood approach where the additional parameters for the missing label mechanism are treated as nuisance parameters. Estimation of the mixture model parameters, $\vect{\Psi}$, in the joint modelling approach is similar to fitting a standard finite mixture model with a nonlinear penalty function. 

We compare our method to fractionally supervised classification, a pseudo-likelihood framework for semi-supervised learning proposed by \citet{Vrbik2015}. \citet{Vrbik2015} also aim to improve on the ignorance likelihood \eqref{eq:ignorance} by assigning different weights to the likelihood contributions of the labelled and unlabelled training observations. A theoretical motiviation is given using weighted likelihood theory \citep{Hu2002}.  Existing empirical results supporting fractionally supervised classification have been obtained under the assumption that class labels are missing completely at random \citep{Vrbik2015, gallaugher_2018_fractionally}. Our theoretical analysis of semi-supervised learning suggests that fractionally supervised classification may not be appropriate when there is a systematic pattern to the missing labels. We examine a number of biomedical datasets with missing labels and find evidence that the realised missing data pattern is related to the difficulty of classification. We present a simulation where the full likelihood estimator based on joint modelling outperforms fractionally supervised classification and the baseline estimator using the ignorance likelihood \eqref{eq:ignorance}.  

\section{Missing data}
\label{sec:missing_data}
\subsection{Overview}
\label{subsec:overview}
Semi-supervised learning falls under the broad umbrella of missing data analysis. \cite{RUBIN1976} provides a useful theoretical framework for conducting statistical inference with missing data. We review some key concepts that are useful for model based semi-supervised learning. The presentation here closely follows the material in  \cite{Little2002} and \cite{Mealli2015}. 

Let $\vect{D}$ denote a $N \times K$ random data matrix. The random matrix $\vect{D}$ models the data before the application of a missingness process. Each element $D_{ij}$ is a scalar random variable for $i=1, \ldots, N$, $j=1, \ldots, K$. Let $\vect{R}$ denote a $N \times K$ random matrix of observation indicators, so $R_{ij}=1$ is $D_{ij}$ is observed in the experiment and $R_{ij}=0$ if $D_{ij}$ is missing. As discussed, the missingness mechanism is an important consideration in statistical inference with missing data. More formally, the missing data mechanism is the conditional distribution of $\vect{R}$ given $\vect{D}$ and a parameter $\vect{\phi}$. Let $\Omega_{\phi}$ denote parameter space of $\phi$. Let $\vect{d}$ represent a matrix of realised data points. The probability that $\vect{R}$ takes the value $\vect{r}$ given that $\vect{D}=\vect{d}$ is denoted $p(\vect{R}=\vect{r} \mid \vect{D}=\vect{d}; \vect{\phi})$. Define the set $I$ as $I=\left\lbrace 1, \ldots, N \right\rbrace \times \left\lbrace 1, \ldots, K \right\rbrace$. The set $I$ contains all possible row-column index tuples for the data matrix $\mat{D}$. The realised value $\vect{r}$ leads to a partition of $I$ into to two disjoint sets, $\text{`mis'}$ and  $\text{`obs'}$. These are defined as $\text{mis}= \left\lbrace (i, j) : r_{ij}=0 \right\rbrace$ and $\text{obs}=\left\lbrace (i, j) : r_{ij}=1 \right\rbrace$. The set $\text{`mis'}$ contains the indices for the observed data and the set $\text{`obs'}$ contains the indices for the missing data. We have a corresponding partition of the data matrix $\vect{D}$ into observed and unobserved components $\vect{D}_{\text{mis}}=(D_{ij} : r_{ij}=0)$ and $\vect{D}_{\text{obs}}=(D_{ij} : r_{ij}=1)$. The sample realised data $\vect{d}$ can also be split into an known component ${\vect{d}}_{\text{obs}}=(d_{ij} : r_{ij}=1)$ and an unknown component $\vect{d}_{\text{mis}}=(d_{ij} : r_{ij}=0)$. 

Suppose we have some parametric model for the data $\vect{D}$. Let $f(\mat{d}_{\text{mis}}, \mat{d}_{\text{obs}} ; \vect{\theta})$ denote the probability mass or density function of the joint distribution of $\mat{D}_{\text{mis}}$ and $\mat{D}_{\text{obs}}$. Given the missingness model, $p(\vect{R}=\vect{r} \mid \vect{D}=\vect{d} ; \vect{\phi})$, the joint density of the observed data, the hidden unobserved data and the missingness indicators can be written as $
  f(\vect{d}_{\text{obs}}, \vect{d}_{\text{mis}}, \vect{r}; \vect{\theta}, \vect{\phi}) = f(\vect{d}_{\text{obs}}, \vect{d}_{\text{mis}} ; \vect{\theta})p(\vect{r} \mid \vect{d}_{\text{mis}}, \vect{d}_{\text{obs}} ; \vect{\phi})$. 
The full likelihood given the observed data, $\vect{d}_{\text{obs}}$, and the missingness indicators ,$\vect{r}$, can be obtained by integrating out the missing observations:
\begin{align}
      L_{\text{full}}(\vect{\theta}, \vect{\phi} ; \vect{d}_{\text{obs}}, \vect{r})     &= \int f(\vect{d}_{\text{obs}}, \vect{d}_{\text{mis}} \mid \vect{\theta})p(\vect{r} \mid \vect{d}_{\text{mis}}, \vect{d}_{\text{obs}} ; \vect{\phi}) \ \text{d}\vect{d}_{\text{mis}}. \label{eq:full_likelihood}
\end{align}
A simpler approach to inference is to ignore the missing data mechanism, and to form the ignorance likelihood 
\begin{align}
    L_{\text{ign}}(\vect{\theta} ; \vect{d}_{\text{obs}}, \vect{r}) &= f(\vect{d}_{\text{obs}} ; \vect{\theta}). \label{eq:ign_likelihood}
\end{align}
We would like to determine when it is appropriate to use the ignorance likelihood \eqref{eq:ign_likelihood} in place of the full likelihood  \eqref{eq:full_likelihood} when $\vect{\theta}$ is the object of interest. There are two theoretical concepts that help to address this question. These are the missing at random principle and the assumption of distinctness. 

\subsection{Missing at random}
A core concept in the analysis of missing data problems is the idea that the missing data are `missing at random' (Definition \ref{defn:mar}).
\begin{definition}[Missing at random.\citep{RUBIN1976}]
\label{defn:mar}
The missing data are said to be missing at random if
\begin{align*}
p(\vect{R}={\vect{r}} \mid \vect{D}_{\emph{obs}}={\vect{d}}_{\emph{obs}}, \vect{D}_{\emph{mis}}=\vect{d}_{\emph{mis}}, \vect{\phi})
\end{align*}
takes the same value for all $\vect{d}_{\emph{mis}}$ and all $\vect{\phi}$. This implies that:
\begin{align*}
p(\vect{R}={\vect{r}} \mid \mat{D}_{\emph{obs}}={\vect{d}}_{\emph{obs}}, \mat{D}_{\emph{mis}}=\vect{d}_{\emph{mis}}, \vect{\phi})  =     p(\vect{R}={\vect{r}} \mid \mat{D}_{\emph{obs}}={\vect{d}}_{\emph{obs}}, \mat{D}_{\emph{mis}}=\vect{d}_{\emph{mis}}', \vect{\phi})
\end{align*}
for all $\vect{d}_{\emph{mis}}$, $\vect{d}_{\emph{mis}}'$ and $\vect{\phi}$.
\end{definition}
The missing at random requirement states the likelihood of the realised missing data pattern $\vect{r}$ is not a function of the unobserved data points. Definition \ref{defn:always_mar} gives a stronger condition that implies Definition \ref{defn:mar}.
\begin{definition}[Missing always at random.  \citep{Mealli2015}] 
\label{defn:always_mar}
The missing data are said to be missing always at random if
\begin{align*}
p(\vect{R}=\vect{r} \mid \vect{D}_{\emph{obs}}=\vect{d}_{\emph{obs}}, \vect{D}_{\emph{mis}}=\vect{d}_{\emph{mis}}, \vect{\phi})=p(\vect{R}=\vect{r} \mid \vect{D}_{\emph{obs}}=\vect{d}_{\emph{obs}}, \vect{D}_{\emph{mis}}=\vect{d}_{\emph{mis}}', \vect{\phi})   
\end{align*}
for all $\vect{r}, \vect{d}_{\emph{obs}}, \vect{d}_{\emph{mis}}, \vect{d}_{\text{mis}}'$, and all $\vect{\phi}$.
\end{definition}
Missing always at random is a stronger requirement, but is perhaps more intuitive. Definition \ref{defn:always_mar} imposes a structural condition on the missing data mechanism. The model for the missing data process needs to satisfy a condition for all possible missing data patterns $\vect{r}$, this is in contrast to Definition \ref{defn:mar} where only the in sample missing data pattern is relevant.  Another particular case of interest is when the missing data mechanism is statistically independent of the realised data. 
\begin{definition}[Missing completely at random. \citep{Mealli2015}]
\label{defn:mcar}
The missing data are missing completely at random if 
\begin{align*}
p(\vect{R}={\vect{r}} \mid \mat{D}=\vect{d}, \vect{\phi})  =     p(\vect{R}={\vect{r}} \mid \vect{\phi})
\end{align*}
for all $\mat{d}$ and $\vect{\phi}$.
\end{definition}
The missing data process has little impact on the analysis when Definition \ref{defn:mcar} is satisfied, however many realistic missingness models will not be compatible with the strict requirements of Definition \ref{defn:mcar} \citep{Little2002}. \citet{seaman_2013_what} provide a detailed discussion on the missing at random assumption and the differences between Definitions \ref{defn:mar}, \ref{defn:always_mar}, and \ref{defn:mcar}.
\subsection{Distinctness}
The distinctness assumption involves the relationship between the parameter of the missingness mechanism and the parameter of the complete-data model. The intuition behind the distinctness assumption that the parameter for the complete-data model, $\vect{\theta}$, does not influence the missingness mechanism. To be more formal, recall that we have defined the parameter space of the model parameter, $\vect{\theta}$, as $\Omega_{\theta}$, and the parameter space of the missingness mechanism parameter, $\vect{\phi}$, as $\Omega_{\phi}$. Let the joint parameter space of the model and missingness parameters be given by $(\vect{\theta}, \vect{\phi}) \in \Omega_{\theta, \phi}$. The distinctness assumption asserts that $\Omega_{\theta, \phi}$ is given by the Cartesian product $\Omega_{\theta} \times \Omega_{\phi}$ \citep{RUBIN1976}. The distinctness assumption leads to a useful simplification of the full likelihood \eqref{eq:full_likelihood}, and is generally treated as a mild condition in applied missing data problems \citep{schafer_1997_analysis}. 

\subsection{Ignorability and efficiency}
 Suppose the data are missing at random, and the distinctness assumption holds. If this is the case, we can establish an equivalence between the full likelihood \eqref{eq:full_likelihood} and the ignorance likelihood \eqref{eq:ign_likelihood}. If the data are missing at random, $p(\vect{r} \mid \vect{d}_{\text{mis}}, \vect{d}_{\text{obs}} ; \vect{\phi}) = p(\vect{r} \mid \vect{d}_{\text{obs}} ; \vect{\phi})$, and the missingness mechanism can be pulled outside the integral:
 \begin{align*}
        L_{\text{full}}(\vect{\theta}, \vect{\phi} ; \vect{d}_{\text{obs}}, \vect{r})&= \int f(\vect{d}_{\text{obs}}, \vect{d}_{\text{mis}} ; \vect{\theta})p(\vect{r} \mid \vect{d}_{\text{obs}} ; \vect{\phi}) \ \text{d}\vect{d}_{\text{mis}} \\
        &= p(\vect{r} \mid \vect{d}_{\text{obs}} ; \vect{\phi})\int f(\vect{d}_{\text{obs}}, \vect{d}_{\text{mis}} ; \vect{\theta})\ \text{d}\vect{d}_{\text{mis}} \\
        &=  p(\vect{r} \mid \vect{d}_{\text{obs}} ; \vect{\phi})f(\vect{d}_{\text{obs}} ; \vect{\theta}).
 \end{align*}
If the distinctness assumption holds, the term $p(\vect{r} \mid \vect{d}_{\text{obs}} ; \vect{\phi})$ is a constant with respect to $\vect{\theta}$. If $\vect{\theta}$ is of primary interest, we can effectively drop $\vect{\phi}$ from the full likelihood and write
\begin{align*}
     L_{\text{full}}(\vect{\theta} ; \vect{d}_{\text{obs}}, \vect{r})
     &\propto f(\vect{d}_{\text{obs}} ; \vect{\theta}).
\end{align*}
Maximum likelihood inference regarding $\vect{\theta}$ using direct application of the ignorance likelihood will lead to the same conclusions as use of the full likelihood \citep{RUBIN1976}.  If the distinctness assumption is violated, the term $p(\vect{r} \mid \vect{d}_{\text{obs}} ; \vect{\phi})$ is no longer a constant with respect to $\vect{\theta}$. The full likelihood then has a contribution from the missing data mechanism, 
\begin{align}
      L_{\text{full}}(\vect{\theta}, \vect{\phi} ; \vect{d}_{\text{obs}}, \vect{r})&=  p(\vect{r} \mid  \vect{d}_{\text{obs}} ; \vect{\phi}, \vect{\theta})f(\vect{d}_{\text{obs}} ; \vect{\theta}). \label{eq:lk_full_efficiency}
\end{align}
It is possible to discard the information in the missingness mechanism and to work with the ignorance likelihood. Integrating over the missingness indicators $\vect{r}$, gives a marginal likelihood in terms of $\vect{\theta}$:
\begin{align}
      L_{\text{marginal}}(\vect{\theta} ; \vect{d}_{\text{obs}}) &=  \int L_{\text{full}}(\vect{\theta}, \vect{\phi} ; \vect{d}_{\text{obs}}, \vect{r})    \ \text{d}\vect{r} \nonumber  \\
      &= f(\vect{d}_{\text{obs}} ; \vect{\theta})\int p(\vect{r} \mid \vect{d}_{\text{obs}} ; \vect{\phi}, \vect{\theta}) \ d\vect{r} \nonumber \\
      &=  f(\vect{d}_{\text{obs}} ; \vect{\theta}). \label{eq:marginal_lk}
\end{align}
If the data are missing at random, but the distinctness assumption does not hold, inference using the ignorance likelihood \eqref{eq:ign_likelihood} will still be valid, but less efficient than the full likelihood approach. The loss in efficiency can be related to the Fisher information in the missingness mechanism $p(\vect{r} \mid \vect{d}_{\text{obs}} ; \vect{\phi}, \vect{\theta})$ \citep{Little2002}. In practice, it is often recommended to use the observed information, as it can be difficult to evaluate the expected information when the data are not missing completely at random \citep{kenward_1998_likelihood}. 
\section{Application to model based classification}
\label{sec:missing_labels}
\subsection{Missing at random}
For the semi-supervised learning problem, the full data matrix $\mat{D}$  consists of the cluster label matrix $\mat{Z}$ and the features $\mat{X}$. Given $n$ records with $p$ dimensional features and $g$ groups, the label indicator matrix is $n \times g$, the matrix $\mat{X}$ is $n \times p$, and the combined data matrix $\mat{D}$ is then $n \times (g+p)$. The full data matrix $\mat{D}$ can be written as 
\begin{align}
    \mat{D} &=  \left[\begin{array}{c|c}  \mat{Z} & \mat{X}\end{array}\right]\\
    &= \left[\begin{array}{cccc|c c c c}
Z_{11} & Z_{12} & \cdots & Z_{1g} & X_{11} & X_{12} & \cdots & X_{1p}\\
Z_{21} & Z_{22} & \cdots & Z_{2g} & X_{21} & X_{22} & \cdots & X_{2p} \\
\vdots & \vdots & \ddots & \vdots & \vdots & \vdots &\ddots &\vdots  \\
Z_{n1} & Z_{n2} & \cdots & Z_{ng} &  X_{n1} & X_{n2} & \cdots & X_{np}\\
\end{array}\right].
\end{align}
We assume that the features $\vect{X}$ are always fully observed and that some labels are possibly missing. For semi-supervised learning problems we introduce a random vector of labelling indicators $\vect{R}=(R_{1}, \ldots, R_{n})^{\T}$. The sample realised missing label pattern is given by $\vect{r}=(r_{1}, \ldots, r_{n})^{\T}$ where observation $i$ is labelled if $r_{i}=1$ and observation is unlabelled if $r_{i}=0$. Observed labels are represented by the $n_{1} \times g $ matrix $\mat{Z}_{\text{obs}}=({Z}_{ih} : r_{i}=1, i \in \set{1, \ldots, n}, h \in \set{1, \ldots, g})$. The missing labels are represented by  the $n_{2} \times g$ matrix $\mat{Z}_{\text{mis}}=({Z}_{ih} : r_{i}=0, i \in \set{1, \ldots, n}, h \in \set{1, \ldots, g})$. The observed data $\mat{D}_{\text{obs}}$ are the $n \times p$ feature matrix $\mat{X}$ and the $n_{1} \times g$ labels indicator matrix $\mat{Z}_{\text{obs}}$. The missing data $\vect{D}_{\text{mis}}$ consists of the $n_{2} \times g$ missing labels $\mat{Z}_{\text{mis}}$.

Using the same notation as in the introduction, the sample realised values, $\mat{d}_{\text{obs}}$, are given by $\vect{x}^{(1)}$, $\vect{z}^{(1)}$ and $\vect{x}^{(2)}$. The sample realised values, $\mat{d}_{\text{mis}}$, are given by $\vect{z}^{(2)}$. The full data likelihood can be written as
\begin{align*}
    L_{\text{full}}(\vect{\Psi}, \vect{\phi} ;\vect{x}^{(1)}, \vect{z}^{(1)}, \vect{x}^{(2)}, \vect{r})  &= \int f(\vect{x}^{(1)}, \vect{z}^{(1)}, \vect{x}^{(2)}, \vect{z}^{(2)}; \vect{\Psi})p(\vect{r} \mid \vect{x}^{(1)}, \vect{z}^{(1)}, \vect{x}^{(2)}, \vect{z}^{(2)}; \vect{\phi}) \ \text{d}\vect{z}^{(2)},
\end{align*}
where $\vect{\Psi}$ represents the set of parameters for the mixture model, and $\vect{\psi}$ is the parameter for the missingness mechanism. We assume the labels are missing at random, so the missingness mechanism is not a function of the unknown labels $\vect{z}^{(2)}$.  The full likelihood then reduces to
\begin{align*}
        L_{\text{full}}(\vect{\Psi}, \vect{\phi} ;\vect{x}^{(1)}, \vect{z}^{(1)}, \vect{x}^{(2)}, \vect{r} )
        &= p(\vect{r} \mid \vect{x}^{(1)}, \vect{z}^{(1)},\vect{x}^{(2)}; \vect{\phi})L_{\text{ign}}(\vect{\Psi} ; \vect{x}^{(1)}, \vect{z}^{(1)}, \vect{x}^{(2)}).
\end{align*}
If the distinctness assumption holds, $p(\vect{r} \mid \vect{x}^{(1)}, \vect{z}^{(1)},\vect{x}^{(2)}; \vect{\phi})$ will be constant with respect to $\vect{\Psi}$.  We can then drop $\vect{\phi}$ from the full likelihood if $\vect{\Psi}$ is the main object of interest. Inference on $\vect{\Psi}$ will be the same using the full likelihood or the ignorance likelihood. There is the proportional relationship:
\begin{align}
     L_{\text{full}}(\vect{\Psi} ;\vect{x}^{(1)}, \vect{z}^{(1)}, \vect{x}^{(2)}, \vect{r}) &= p(\vect{r} \mid \vect{x}^{(1)}, \vect{z}^{(1)},\vect{x}^{(2)}; \vect{\phi})L_{\text{ign}}(\vect{\Psi} ; \vect{x}^{(1)}, \vect{z}^{(1)}, \vect{x}^{(2)}) \nonumber \\
     & \propto L_{\text{ign}}(\vect{\Psi} ; \vect{x}^{(1)}, \vect{z}^{(1)}, \vect{x}^{(2)}). \label{eq:distinctness_relation}
\end{align}
When the labels are missing at random and the distinctness assumption holds, the pattern of missing labels is ignorable. Under these conditions, maximum likelihood inference on $\vect{\Psi}$ using the ignorance likelihood will be equivalent to maximum likelihood inference using the full data likelihood. Now suppose that the distinctness assumptions does not hold, so $\vect{\Psi}$ enters the missingness mechanism in some manner. For convenience, assume that we can represent the parameter of the missingness mechanism as $\vect{\phi}=(\vect{\Psi}, \vect{\beta})$, where $\vect{\beta}$ is a supplementary parameter.  In this case, the full semi-supervised likelihood has a contribution from the missingness mechanism, and a contribution from the observed data:
\begin{align}
     L_{\text{full}}(\vect{\Psi}, \vect{\beta} ;\vect{x}^{(1)}, \vect{z}^{(1)}, \vect{x}^{(2)}, \vect{r}) &= p(\vect{r} \mid \vect{x}^{(1)}, \vect{z}^{(1)},\vect{x}^{(2)}; \vect{\Psi}, \vect{\beta})L_{\text{ign}}(\vect{\Psi} ; \vect{x}^{(1)}, \vect{z}^{(1)}, \vect{x}^{(2)}). \label{eq:full_semi_lk}
\end{align}
If the distinctness assumption does not hold, there is information about $\vect{\Psi}$ in the observed data $(\vect{x}^{(1)}, \vect{z}^{(1)}, \vect{x}^{(2)})$ and the missingness pattern $\vect{r}$. Joint modelling is required for fully efficient estimation of $\vect{\Psi}$. The estimator using the ignorance likelihood is still consistent, but suboptimal. If the Fisher information provided by the missingness mechanism is high, we can expect the full likelihood estimator to outperform the ignorance likelihood estimator by a wide margin. 

\section{Missingness mechanisms}
\label{sec:missingness}
\subsection{Distinctness}
As mentioned in the introduction, in many practical applications cluster labels will be assigned by experts. Manual annotation of the dataset can induce a systematic missingness mechanism. Our main idea is that the probability that a particular observation is unlabelled is related to the difficulty of classifying the observation. As an example, suppose medical professionals are asked to classify each image from a set of MRI scans into three groups, tumour present, no tumour present, or unknown. It seems reasonable to expect that the unknown observations will correspond to images that do not present clear evidence for the presence or absence of a tumour. The unlabelled observations will exist in regions of feature space where there is class overlap.  We will argue that in these situations, the unlabelled observations carry additional information that can be used to improve the efficiency of parameter estimation. 

As mentioned in the Introduction,the difficulty of classifying an observation  can be measured using the Shannon entropy. Let $e_{i}$ denoted the Shannon entropy for observation $i$, as in \eqref{eq:shannon_entropy_intro}. Let $\vect{R}=(R_{i}, \ldots, R_{n})^{\T}$ be a random vector of labelling indicators for the data. Observation $i$ is labelled if $R_{i}=1$, and observation $i$ is unlabelled if $R_{i}=0$ for $i=1, \ldots, n$. We propose to model the log odds of an observation being labelled as a function of the entropy. For flexible modelling, suppose we have $T$ basis functions $b_{1}, \ldots, b_{T}$, where $b_{t}:\mathbb{R} \to \mathbb{R}$. Let $\beta_{0}, \beta_{1}, \ldots, \beta_{T}$ denote scalar coefficients, and $\vect{\beta}=(\beta_{0}, \ldots, \beta_{T})^{\T}$. We assume that for each observation $i=1, \ldots, n$:
\begin{align}
     \log \dfrac{\Pr(R_{i}=1 \mid \vect{x}_{i} ; \vect{\Psi}, \vect{\beta})}{\Pr(R_{i}=0 \mid \vect{x}_{i} ; \vect{\Psi}, \vect{\beta})} &= \beta_{0}+\sum_{t=1}^{T}\beta_{t}b_{t}(e_{i}).  \label{eq:general_log_odds_entropy}
\end{align}
If the missing label mechanism is of the form \eqref{eq:general_log_odds_entropy}, the distinctness assumption will not hold as $e_{i}$ is a function of the mixture model parameter $\vect{\Psi}$, entering through \eqref{eq:posterior_class_intro}. The full likelihood with respect to $\vect{\Psi}$ involves a contribution from the missingness mechanism:
\begin{align*}
     L_{\text{full}}(\vect{\Psi}, \vect{\beta} ; \vect{r}, \vect{x}^{(1)}, \vect{z}^{(1)}, \vect{x}^{(2)}) &= p(\vect{r} \mid \vect{x}^{(1)}, \vect{x}^{(2)}; \vect{\Psi}, \vect{\beta})L_{\text{ign}}(\vect{\Psi}; \vect{x}^{(1)}, \vect{z}^{(1)}, \vect{x}^{(2)}).
\end{align*}
Maximum likelihood inference ignoring the missing data mechanism will be less efficient than an approach that models the labelling probability explicitly. To summarise, the pattern of missing labels can provide information about $\vect{\Psi}$ when the missingness mechanism is related to the Shannon entropy of the posterior class probabilities. The parametric model defines regions of high class uncertainty, and we wish for the location of the unlabelled observations to be consistent with the model. Before proceeding any further, we examine some real datasets to test the hypothesis that the missingness mechanism can be related to the Shannon entropy. For visualisation we find it useful to transform the Shannon entropy values to the real line. The entropy $e_{i}$ can be mapped to the unit interval through $e_{i}/\log(g)$, application of the logit transformation then gives the transformed entropy $e'_{i}=\log[\{e_{i}/\log(g)\}/\{1-e_{i}/\log(g)\}]$ for $i=1, \ldots, n$.

\subsection{Flow cytometry dataset}
\begin{figure}
\centering
\includegraphics[width=0.8\textwidth]{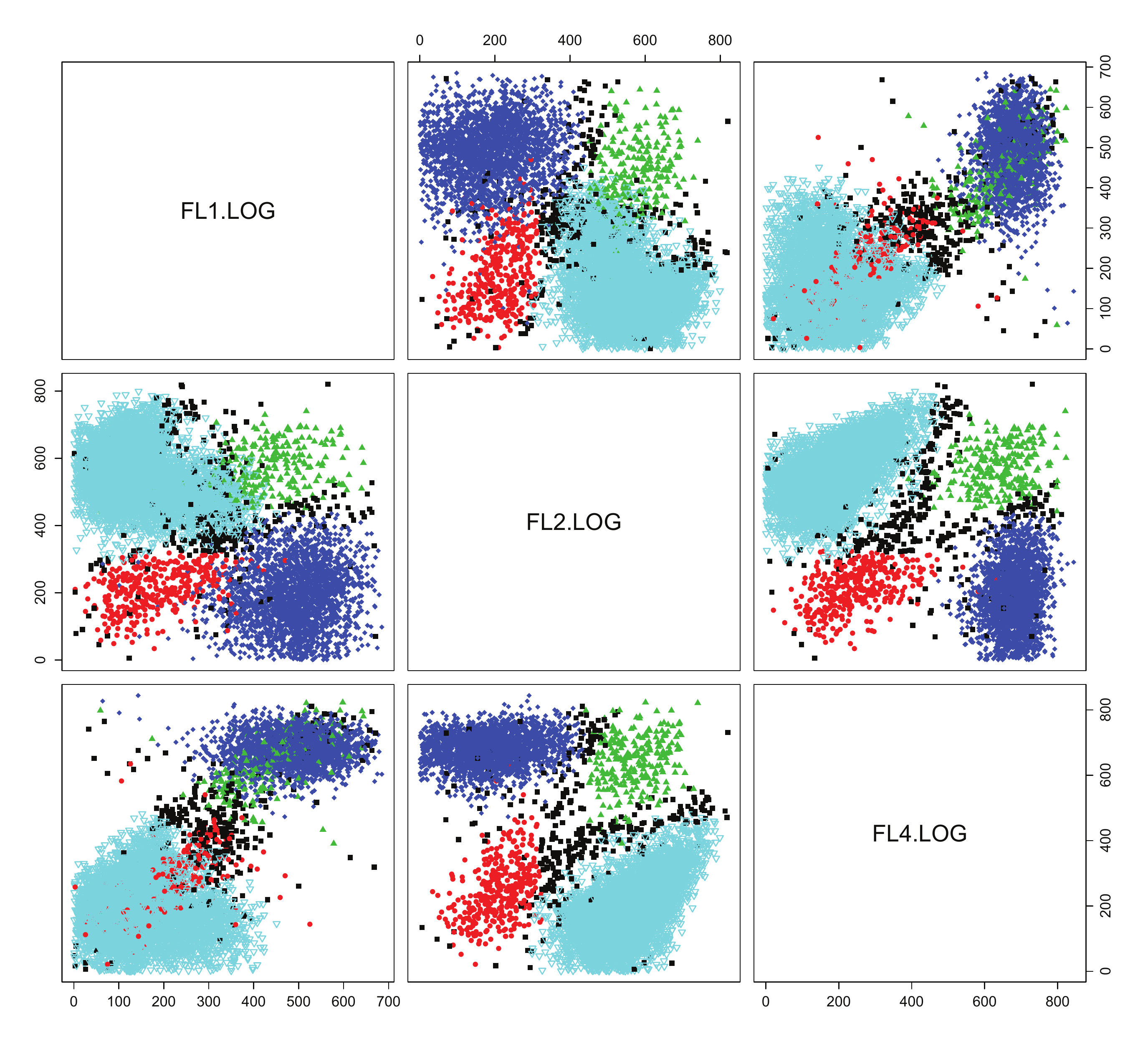}
\caption{Flow cytometry dataset with manually assigned labels. Black squares denote observations that were unlabelled by the expert.}
    \label{fig:cytometry_data}
\end{figure}

We first consider a flow cytometry dataset from \citet{Aghaeepour2013}. The dataset consists of fluorescence measurements on $n=11,792$ cells using $p=3$ markers. Cluster labels were assigned manually by domain experts using specialised software for the analysis of flow cytometry data. Labels were assigned using a combination of user defined `gates' that partition feature space into groups. There were $n_{2}=333$ observations that were not assigned to a group at the end of the manual gating process. Figure \ref{fig:cytometry_data} shows a pairs plot of the dataset. Black squares denote unlabelled observations. Clusters are plotted using different colours and shapes. The unlabelled observations appear to be in areas where there is some overlap between clusters. Unlabelled observations appear to be concentrated around class decision boundaries, supporting the general idea that experts will hesitate to label observations that are difficult to classify. 

\begin{figure}
\includegraphics[width=\textwidth]{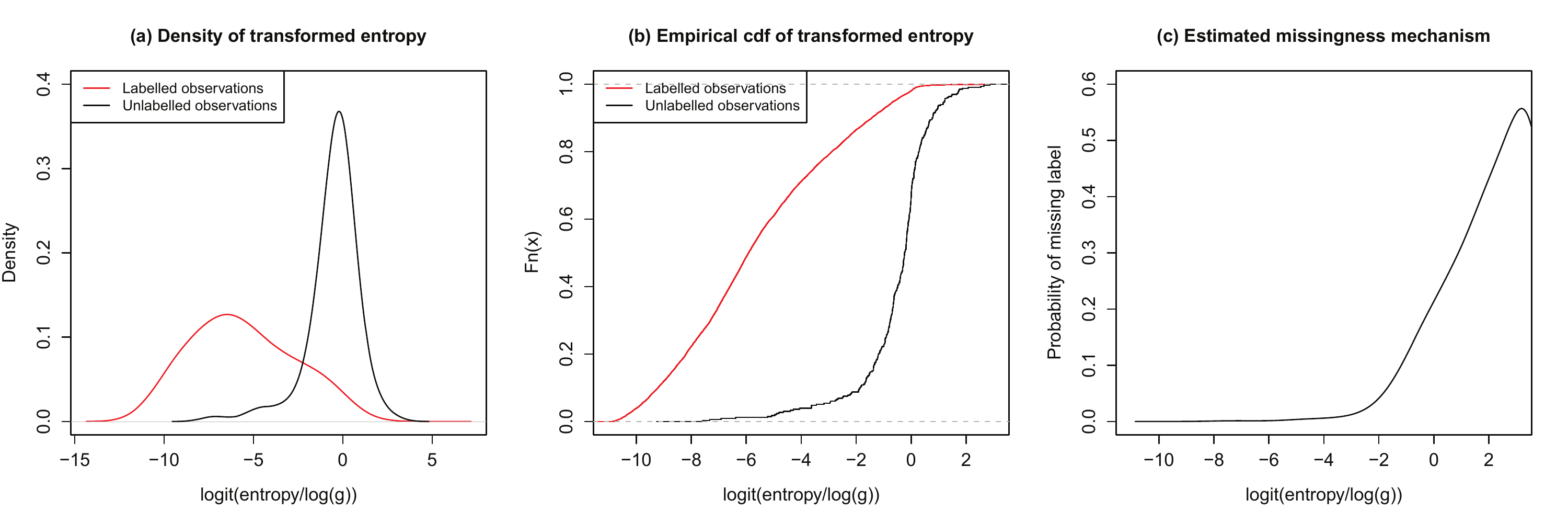}
\caption{Analysis of cytometry dataset. Panel (a) compares the transformed entropy of the labelled and unlabelled observations using kernel density estimates. Panel (b) compares the empirical cumulative distribution function of the unlabelled and unlabelled observations Unlabelled observations have higher entropy than the labelled observations. Panel (c) shows a nonparametric estimate of the missing data mechanism. The probability of labelling appears to be a smooth function of transformed entropy.}
    \label{fig:cytometry_analysis}
\end{figure}

We fit a skew-\textit{t} mixture model to estimate the classification entropy of each observation.  Figure \ref{fig:cytometry_analysis} (a) compares kernel density estimates of the transformed entropy of the labelled and unlabelled observations. Panel (b) compares the empirical cumulative distribution functions of the estimated transformed entropy distributions in the labelled and unlabelled groups. Panel (c) shows a Nadaraya-Watson kernel estimate of the labelling probability. From (a) and (b), we can see that the unlabelled observations typically have higher entropy than the labelled observations. The estimated missing label probability in (c) appears to be a smooth function of the transformed entropy.

\subsection{Cardiotocography dataset}
\begin{figure}
\centering
\includegraphics[width=0.5\textwidth]{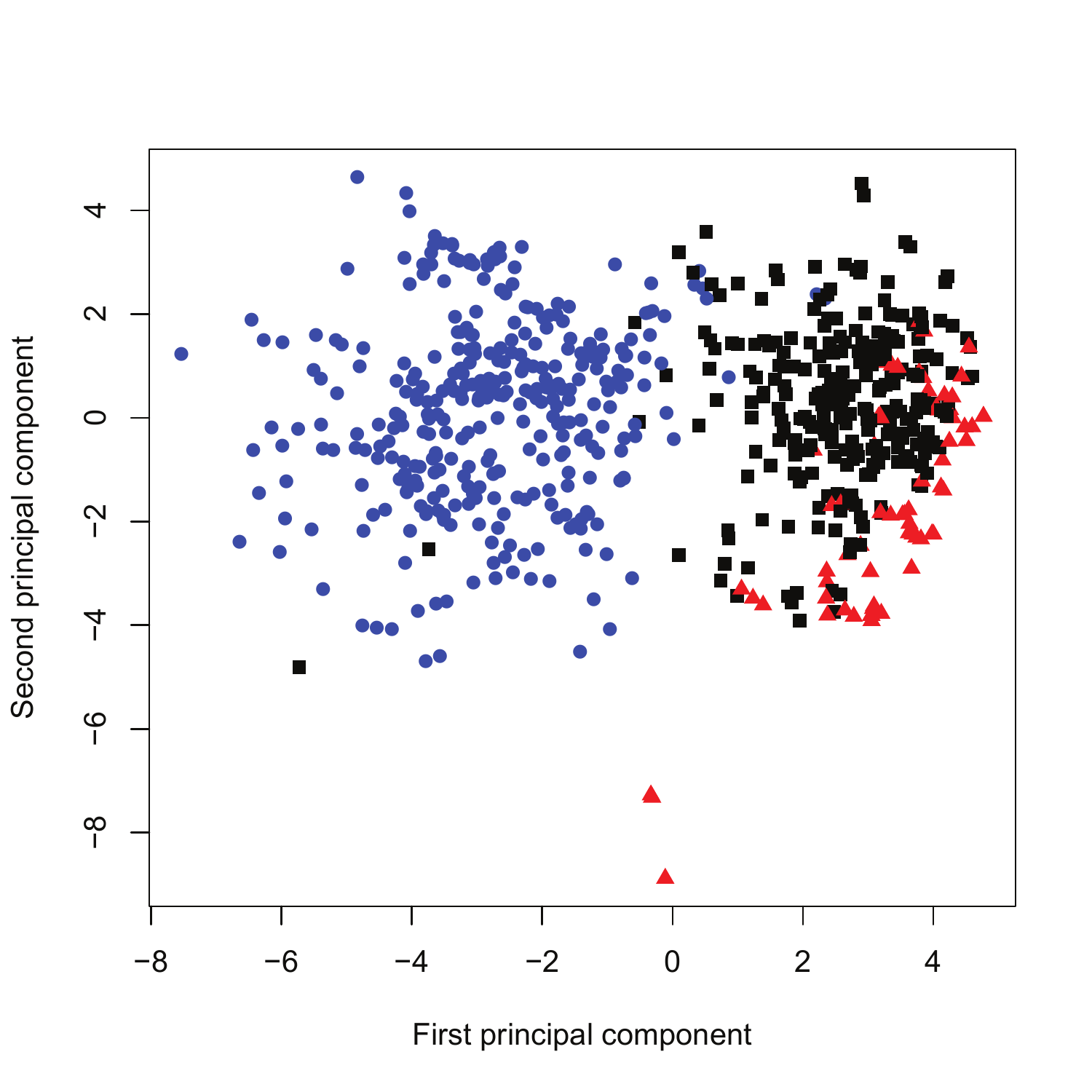}
\caption{Cardiotocography dataset. Blue labels and red triangles are for normal and pathological observations respectively. Black squares correspond to unlabelled observations.}
\label{fig:cardio_data}
\end{figure}

We consider a subset of data from \citet{Ayres-de-Campos2000}. The full dataset consists of 23 features extracted from cardiotocograms on 2126 infants.  A panel of three obstetricians  used the cardiotocograms to assess fetal state. Observations were labelled as normal, pathological or suspect given the expert consensus. We take the suspect observations to be unlabelled. The cardiograms were also assigned a morphological pattern $(1,\ldots, 10)$ using automated methods. We restricted attention to the observations with morphological patterns $5,6,9$ and $10$ as the majority of the unlabelled observations are in these groups. The subset we considered has $n=670$ observations, with $n_1=402$ labelled observations and $n_{2}=268$ unlabelled observations. We performed dimension reduction using principal components analysis prior to clustering. We worked with the first two principal component scores.  Figure \ref{fig:cardio_data} shows the data subset. Normal observations are plotted as blue circles, pathological observations are plotted as red triangles. Suspect observations are plotted as black squares. The bulk of the unlabelled observations are concentrated between the normal and pathological groups. Unlabelled observations appear to be in regions where there is class uncertainty.

\begin{figure}
\includegraphics[width=\textwidth]{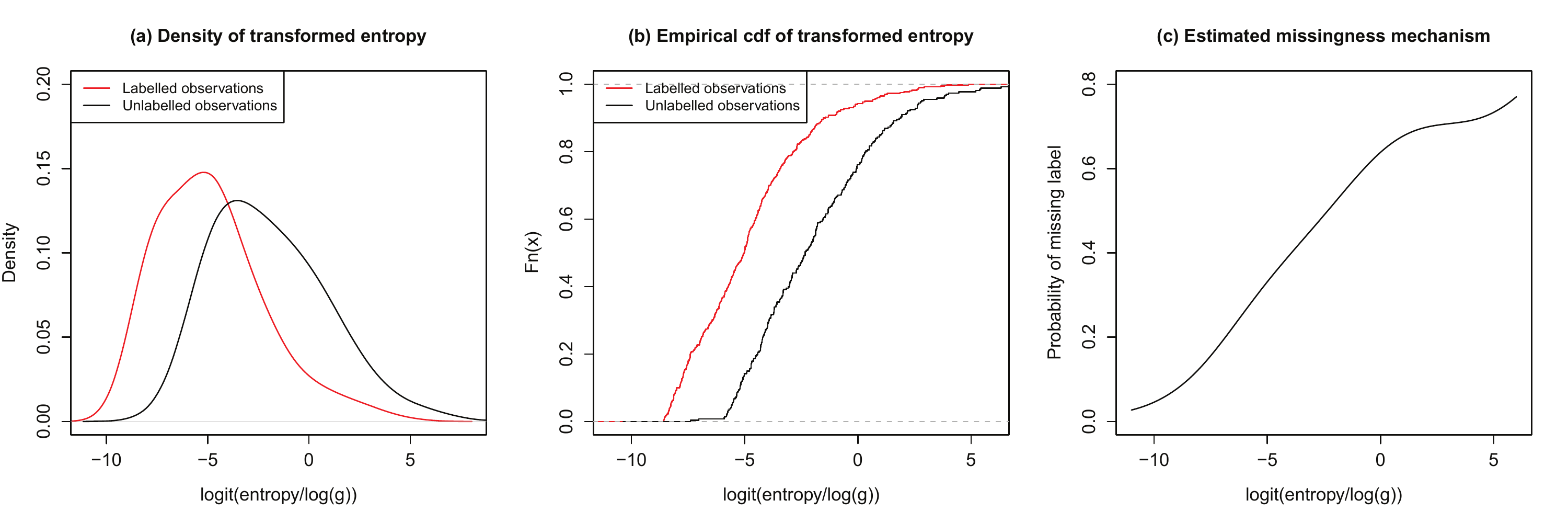}
    \caption{Analysis of cardiotocography dataset. Panel (a) compares the entropy of the labelled and unlabelled observations using kernel density estimates.  Panel (b) compares the ecdf of the unlabelled and unlabelled observations.  Unlabelled observations have higher Shannon entropy than the labelled observations. Panel (c) shows a nonparametric estimate of the missing data mechanism. The probability of labelling appears to be a smooth function of entropy.}
    \label{fig:cardio_analysis}
\end{figure}

We fit a two-component skew-\textit{t} mixture model to the observed dataset. We then estimated the entropy of each observation. Figure \ref{fig:cardio_analysis} compares the transformed entropy of the labelled and unlabelled observations. Panel (a) compares kernel density estimates and panel (b) compares the empirical cumulative distribution functions. Panel (c) shows a Nadaraya-Watson kernel estimate of the labelling probability. From (a) and (b), we can see that the unlabelled observations typically have higher entropy than the labelled observations. The estimated missing label probability in (c) appears to be a smooth function of the transformed entropy.

\subsection{Gastrointestinal dataset}
We also consider a subset of data from \citet{Mesejo2016}. The raw dataset consists of 700 features extracted from colonoscopic videos on patients with gastrointestinal lesions. There are $n=152$ records. A panel of seven doctors reviewed the videos and determined whether the lesions appeared benign or malignant. We formed a consensus labelling using the individual expert labels. Observations where six or more of the experts agreed were treated as labelled. Observations where fewer than six experts agreed were treated as unlabelled. 

\begin{figure}
\includegraphics[width=\textwidth]{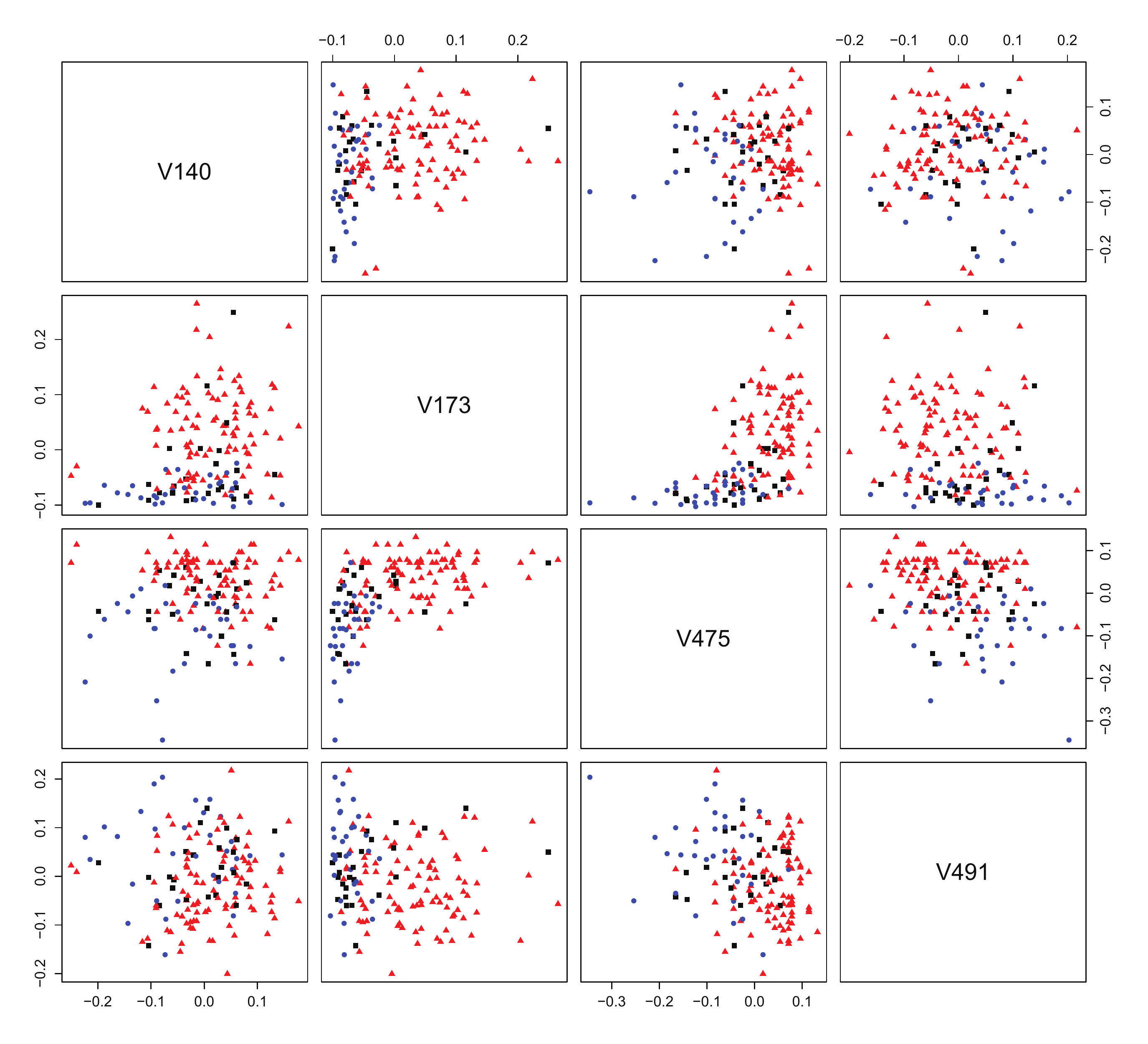}
    \caption{Gastrointestinal dataset. Red triangles denote benign observations and blue circles denote malignant observations. Black squares correspond to unlabelled observations. Observations are treated unlabelled  if the fewer than 6/7 experts assigned the same class label to the observation.}
    \label{fig:gastro_data}
\end{figure}

The dataset also includes a ground truth set of labels, obtained using additional histological measurements. The accuracy of the experts can be determined by comparing to the ground truth labels. To reduce the dimension of the dataset, we used sparse linear discriminant analysis \citep{clemmensen_2011_sparse}, to select a subset of four features useful for class discrimination using the ground truth labels. These four variables were taken as the features for model based clustering. Figure \ref{fig:gastro_data} shows the data subset. Black squares denote unlabelled observations, red triangles denote benign observations and blue circles denote malignant observations. It seems that the unlabelled observations are located in regions where there is group overlap. This dataset is smaller than the cytometry and cardiotocopgraphy datasets, so the pattern of missingness is less visually distinctive. 

\begin{figure}
\includegraphics[width=\textwidth]{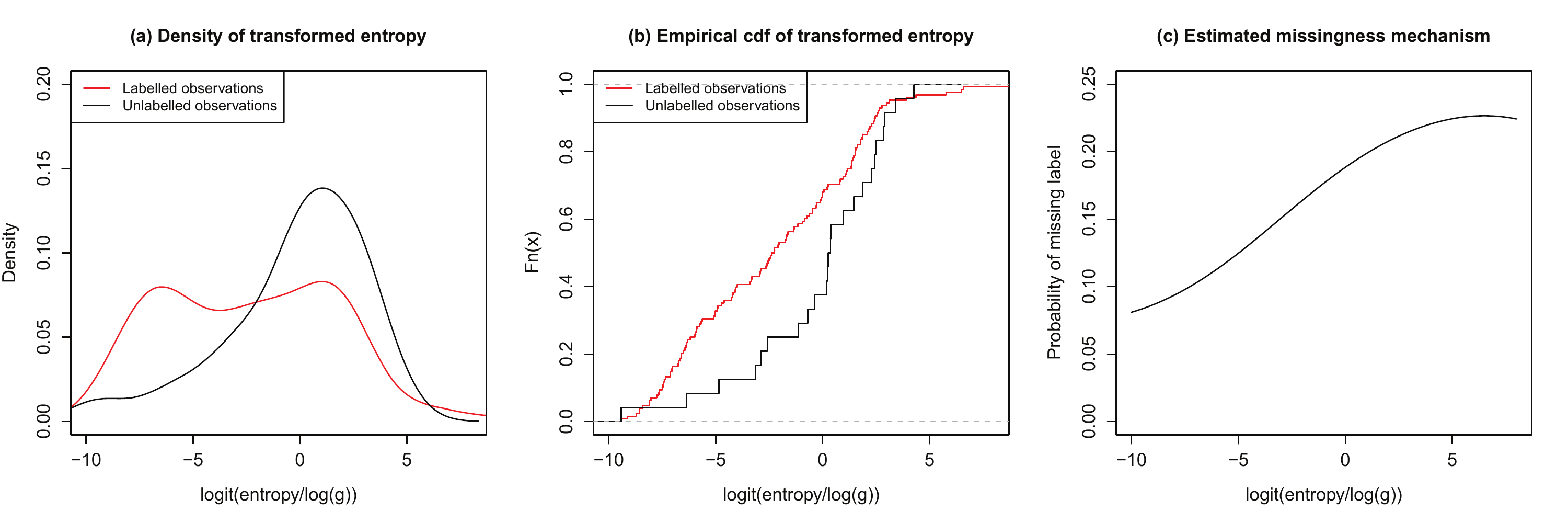}
    \caption{Analysis of gastrointestinal dataset. Panel (a) compares the entropy of the labelled and unlabelled observations using kernel density estimates. Panel (b) compares the ecdf of the unlabelled and unlabelled observations.   Unlabelled observations have higher Shannon entropy than the labelled observations. Panel (c) shows a nonparametric estimate of the missing data mechanism. The probability of labelling appears to be a smooth function of entropy.}
    \label{fig:gastro_analysis}
\end{figure}

We fit a two-component student-\textit{t} mixture model to the dataset. We then used the fitted model to estimate the entropy of each observation. Figure \ref{fig:gastro_analysis} compares the transformed entropy of the labelled and unlabelled observations. Panel (a) compares kernel density estimates and panel (b) compares the empirical cumulative distribution functions. Panel (c) shows a Nadaraya-Watson kernel estimate of the labelling probability. From (a) and (b), we can see that the unlabelled observations typically have higher entropy than the labelled observations. The estimated missing label probability in (c) appears to be a smooth function of the transformed entropy.

\subsection{Summary}
We hypothesised that unlabelled observations will have higher classification Shannon entropy than labelled observations. We examined three datasets with missing labels. Table \ref{tab:missing_summary} gives useful summary statistics for each dataset. The columns $\overline{\mu}_{L}$ and $\overline{\mu}_{U}$ give the average transformed entropy of the labelled and unlabelled observations respectively. In each analysis we found that the unlabelled observations typically had greater Shannon entropy than the labelled observations. 

\begin{table}[]
    \centering
    \begin{tabular}{@{}lllllllll@{}}
    \toprule
    Dataset & $n$ & $p$& $g$ & $n_{1}$& $n_{2}$& $\overline{\mu}_{L}$ & $\overline{\mu}_{U}$ \\ \midrule
    Cytometry & 11,792& 3 & 4 &11,459 & 333 &  -5.59 & -0.53\\  
    Cardiotocography & 669 & 2 & 2 & 401 &268  & -4.82 & -1.92 \\  
    Gastrointestinal & 152 & 4 & 2 &128 &24 & -2.34 & -0.21 \\  \bottomrule
    \end{tabular}
    \caption{Summary statistics for dataset examples. The columns $\overline{\mu}_{L}$ and $\overline{\mu}_{U}$ give the average transformed entropy of the labelled and unlabelled observations respectively. The transformed entropy for observation $i$ is given by $e'_{i}=\log[\{e_{i}/\log(g)\}/\{1-e_{i}/\log(g)\}]$. In each dataset, the unlabelled observations have a higher average transformed entropy than the labelled observations.}
    \label{tab:missing_summary}
\end{table}

We performed statistical hypothesis tests to assess the evidence for a systematic missing label process. We used a one-sided Kolmogorov-Smirnov test using the empirical cumulative distribution functions of the transformed entropy distributions in each dataset. Let $V_{L}$ denote the transformed entropy of a randomly selected labelled  observation. Let $V_{U}$ denote the transformed entropy of a randomly selected unlabelled observation. Let $F_{V_{L}}(v)$ denote the cumulative distribution function for the distribution of $V_{L}$ and let  $F_{V_{U}}(v)$ denote the cumulative distribution function of the distribution of $V_{U}$. The null, $H_{0}$, and alternative, $H_{1}$, are given respectively by:
\begin{align*}
    H_{0} &: F_{V_{L}}(v)  \le F_{V_{U}}(v) \quad \text{for all $v$}, \\
    H_{1} &: F_{V_{L}}(v) > F_{V_{U}}(v)  \quad \text{for at least one $v$}.
\end{align*}
We also performed a one-sided Mann-Whitney $U$ test using the estimated entropy scores. The null, $H_{0}$,  and the alternative, $H_{1}$, are given respectively by:
\begin{align*}
H_{0} &: \Pr(V_{U}   > V_{L}) \le 0.5, \\
H_{1} &: \Pr(V_{U}   > V_{L}) > 0.5.
\end{align*}
Table \ref{tab:hyp_tests} reports the test statistics and $p$-values. We reject the null hypothesis in each test using the 5 percent level of significance. The $p$-values are very small for the cytometry and cardiotocography datasets. In each analysis we find evidence for a systematic missingness mechanism related to classification difficulty. 

\begin{table}[]
\centering
\begin{tabular}{@{}lllll@{}}
\toprule
 & \multicolumn{2}{c}{Kolmogorov-Smirnov test} & \multicolumn{2}{c}{Mann-Whitney U test} \\ \cmidrule(lr){2-3}\cmidrule(lr){4-5}
Dataset & Statistic     & $p$-value     & Statistic     & $p$-value    \\ \midrule
Cytometry &   0.783   & $<  10^{-16}$            &    232996           &    $< 10^{-16}$          \\
Cardiotocography &   0.429         &    $< 10^{-16}$           &  23311             &  $< 10^{-16}$         \\
Gastrointestinal &  0.336           &   $0.0105$           &     1014          &   0.0042        \\ \bottomrule
\end{tabular}
\caption{Hypothesis tests for the data examples. We test the idea that the entropy of the unlabelled observations is higher than the entropy of the labelled observations. We performed one-sided Kolmogorov-Smirnov tests and one-sided Mann-Whitney U tests. We reject the null hypothesis in each test at the 5\% level of significance. There appears to be evidence that there is a statistical relationship between the classification Shannon entropy and the labelling probability.}
\label{tab:hyp_tests}
\end{table}

\section{Estimation}
We take the missing label model to be of the form \eqref{eq:general_log_odds_entropy}. We assume the $T$ basis functions $b_{1}, \ldots, b_{T}$, are known, and that the coefficients $\vect{\beta}=(\beta_{0}, \beta_{1}, \ldots, {\beta}_{T})^{\T}$ must be estimated from the data. Define the posterior class probabilities for the labelled and unlabelled observations as
\begin{align*}
    \tau_{jh}^{(1)} &= \dfrac{\pi_{h}f(\vect{x}_{j}^{(1)} ; \vect{\theta}_{h})}{\sum_{r=1}^{g}\pi_{r}f(\vect{x}_{j}^{(1)} ; \vect{\theta}_{r})} ,  \quad j=1, \ldots, n_{1}; \ h=1, \ldots, g. \\
    \tau_{kh}^{(2)} &= \dfrac{\pi_{h}f(\vect{x}_{k}^{(2)} ; \vect{\theta}_{h})}{\sum_{r=1}^{g}\pi_{r}f(\vect{x}_{k}^{(2)} ; \vect{\theta}_{r})} , \quad k=1, \ldots, n_{2}; \  h=1, \ldots, g.
\end{align*}
Define the classification entropy of the labelled and unlabelled observations as
\begin{align*}
    e_{j}^{(1)} &= -\sum_{h=1}^{g}\tau_{jh}^{(1)} \log \tau_{jh}^{(1)} ,  \quad j=1, \ldots, n_{1}. \\
    e_{k}^{(2)} &= -\sum_{h=1}^{g}\tau_{kh}^{(2)} \log \tau_{kh}^{(2)}, \quad k=1, \ldots, n_{2}.
\end{align*}
Let $\eta_{j}^{(1)}$ denote the log odds of the labelling probability for the $j$th labelled observation and let $\eta_{k}^{(2)}$ denote the log odds of the labelling probability for the $k$th unlabelled observation.
\begin{align}
     \eta_{j}^{(1)} &= \beta_{0}+\sum_{t=1}^{T}\beta_{t}b_{t}(e_{j}^{(1)}), \quad j=1, \ldots, n_{1}. \label{eq:eta_labelled} \\
     \eta_{k}^{(2)} &= \beta_{0}+\sum_{t=1}^{T}\beta_{t}b_{t}(e_{k}^{(2)}), \quad k=1, \ldots, n_{2}. \label{eq:eta_unlabelled}
\end{align}
The full likelihood including the missingness mechanism is then:
\begin{align}
    L_{\text{full}}(\vect{\Psi}, \vect{\beta} ; \vect{r}, \vect{x}^{(1)}, \vect{z}^{(1)}, \vect{x}^{(2)}) &= p(\vect{r} \mid \vect{x}^{(1)}, \vect{x}^{(2)}; \vect{\Psi}, \vect{\beta}) L_{\text{ign}}(\vect{\Psi}; \vect{x}^{(1)}, \vect{z}^{(1)}, \vect{x}^{(2)}) \nonumber \\
    &= \left( \prod_{j=1}^{n_{1}} \dfrac{\exp(\eta_{j}^{(1)})}{1+\exp(\eta_{j}^{(1)})} \right)\left(\prod_{k=1}^{n_{2}} \dfrac{1}{1+\exp(\eta_{k}^{(2)})} \right)  L_{\text{ign}}(\vect{\Psi}; \vect{x}^{(1)}, \vect{z}^{(1)}, \vect{x}^{(2)}). \label{eq:likelihood_with_mechanism} 
\end{align}
The coefficients  $\vect{\beta}=(\beta_{0}, \beta_{1}, \ldots, \beta_{T})^{\T}$ can be treated as nuisance parameters if the primary goal is classification or clustering. Given the mixture parameters $\vect{\Psi}$, the likelihood contribution from the missingness mechanism is of the same general form as a logistic regression model. The missingness indicators take the role of the response in the regression model, and the features are given by $b_{1}(e_{j}^{(1)}), \ldots, b_{T}(e_{j}^{(1)})$ and   $b_{1}(e_{k}^{(2)}), \ldots, b_{T}(e_{k}^{(2)})$ for $j=1, \ldots, n_{1}; k=1, \ldots, n_{2}$. Let $\widetilde{\vect{y}}$ and $\widetilde{\mat{X}}$ give the implicit response vector and design matrix that we use for the missingness mechanism. The display below shows the structure of $\widetilde{\vect{y}}$ and $\widetilde{\mat{X}}$ in detail. The first $n_{1}$ elements of $\widetilde{\vect{y}}$ correspond to the labelled observations and the following $n_{2}$ elements correspond to the unlabelled observations. The first column in $\widetilde{\mat{X}}$ is for the intercept $\beta_{0}$, and the remaining columns are for the coefficients ${\beta}_{1}, \ldots, \beta_{T}$.
\begin{align*}
\widetilde{\vect{y}}=\begin{pmatrix}
1 \\
1 \\
\vdots \\
1 \\ 
0 \\
0 \\
\vdots \\
0
\end{pmatrix}, \quad \widetilde{\mat{X}}=\begin{pmatrix}
1 & b_{1}(e_{1}^{(1)}) & \hdots & b_{T}(e_{1}^{(1)}) \\
1 & b_{1}(e_{2}^{(1)}) & \hdots & b_{T}(e_{2}^{(1)}) \\
\vdots & \vdots & \hdots & \vdots  \\
1 & b_{1}(e_{n_{1}}^{(1)}) & \hdots & b_{T}(e_{n_{1}}^{(1)}) \\
1 & b_{1}(e_{1}^{(2)}) & \hdots & b_{T}(e_{1}^{(2)}) \\
1 & b_{1}(e_{2}^{(2)}) & \hdots & b_{T}(e_{2}^{(2)}) \\
\vdots & \vdots & \hdots & \vdots  \\
1 & b_{1}(e_{n_{2}}^{(2)}) & \hdots & b_{T}(e_{n_{2}}^{(2)}) \\
\end{pmatrix}.
\end{align*}

Let $\widehat{\vect{\beta}}=(\widehat{\beta}_{0}, \widehat{\beta}_{1}, \ldots, \widehat{\beta}_{T})^{\T}$ denote the maximum likelihood estimates of the coefficients conditional on $\vect{\Psi}$. The estimates $\widehat{\vect{\beta}}$ can be obtained using any standard routine for logistic regression using $\widetilde{\vect{y}}$ and $\widetilde{\mat{X}}$. Given $\widehat{\vect{\beta}}$, the fitted linear predictors, $\widehat{\eta}_{j}^{(1)}$ and $\widehat{\eta}_{k}^{(2)}$, can be obtained using \eqref{eq:eta_labelled} and \eqref{eq:eta_unlabelled} for $j=1, 
\ldots, n_{1};k=1, \ldots, n_{2}$. It is then possible to work with the profile likelihood, which is a sole function of the mixture parameters $\vect{\Psi}$,
\begin{align*}
     L_{\text{profile}}(\vect{\Psi}; \vect{r}, \vect{x}^{(1)}, \vect{z}^{(1)}, \vect{x}^{(2)}) &= \left( \prod_{j=1}^{n_{1}} \dfrac{\exp(\widehat{\eta}_{j}^{(1)})}{1+\exp(\widehat{\eta}_{j}^{(1)})} \right)\left( \prod_{k=1}^{n_{2}} \dfrac{1}{1+\exp(\widehat{\eta}_{k}^{(2)})} \right)  L_{\text{ign}}(\vect{\Psi}; \vect{x}^{(1)}, \vect{z}^{(1)}, \vect{x}^{(2)}).
\end{align*}
In the simulations we maximised the likelihood \eqref{eq:likelihood_with_mechanism} using the \texttt{BFGS} method as implemented in the \texttt{optim} function in the \texttt{R} software environment. The one-step late EM algorithm \citep{green1990use} could also be used, treating the missingness mechanism as a penalty function. 

\section{Fractionally supervised classification}
\label{sec:fsc}
\subsection{Overview}
Fractionally supervised classification belongs to the general family of model based classification techniques. 
Ignoring the missing data mechanism, the likelihood function for the labelled data, unlabelled data and the observed data are given by \eqref{eq:lk_labelled} and \eqref{eq:lk_unlabelled} respectively
\begin{align}
    \mathcal{L}_{\text{ign}}^{(1)}(\vect{\Psi}; \vect{x}^{(1)}, \vect{z}^{(1)})  &= \prod_{j=1}^{n_{1}}\prod_{h=1}^{g}[\pi_{h}f(\vect{x}^{(1)}_{j} ; \vect{\theta}_{h})]^{z^{(1)}_{jh}}, \label{eq:lk_labelled} \\
    \mathcal{L}_{\text{ign}}^{(2)}(\vect{\Psi}; \vect{x}^{(2)})  &= \prod_{k=1}^{n_{2}}\sum_{h=1}^{g}[\pi_{h}f(\vect{x}^{(2)}_{k} ; \vect{\theta}_{h})], \label{eq:lk_unlabelled} 
\end{align}
The likelihood factor $\mathcal{L}_{\text{ign}}^{(1)}(\cdot)$ is formed using the labelled data  $(\vect{x}^{(1)}, \vect{z}^{(1)})$, and is reminiscent of the likelihood used in discriminant analysis with no missing data. The likelihood factor $\mathcal{L}_{\text{ign}}^{(2)}(\cdot)$ is formed using the unlabelled data $\vect{x}^{(2)}$ and is reminiscent of the likelihood used in a model based clustering analysis with no missing data. \citet{Vrbik2015} propose to introduce a weight $\alpha \in [0,1]$ to form a pseudo-likelihood  from the labelled and unlabelled likelihood contributions \eqref{eq:lk_labelled} and \eqref{eq:lk_unlabelled}. The objective function for fractionally supervised classification is defined as
\begin{align}
    \mathcal{L}_{\text{FSC}}(\vect{\Psi} \mid \alpha) &= [\mathcal{L}_{\text{ign}}^{(1)}(
    \vect{\Psi}; \vect{x}^{(1)}, \vect{z}^{(1)})]^{\alpha}[ \mathcal{L}_{\text{ign}}^{(2)}(\vect{\Psi} ; \vect{x}^{(2)})]^{(1-\alpha)}. \label{eq:fsc_objective}
\end{align}
The pseudo-likelihood in \eqref{eq:fsc_objective} gives different weights to the unlabelled and labelled training data. \citeauthor{Vrbik2015} view the method as smoothly interpolating between a  fully supervised analysis (discriminant analysis) and a fully unsupervised analysis (cluster analysis). Positive simulation results supporting fractional supervised classification have been obtained under the assumption that the labels are missing completely at random. It is of interest to determine how the procedure behaves when this assumption is relaxed \citep{Vrbik2015, gallaugher_2018_fractionally}. Missing data theory is useful for this purpose, and gives a different perspective than weighted likelihood theory \citep{Hu2002}.  

\subsection{Theoretical properties}
The full likelihood can also be expressed as the product of two conditional likelihoods, taking into account the missingness mechanism. Let $f(\vect{x}_{i} \mid \vect{Z}_{i}=\vect{z}_{i}, R_{i}=1 ; \vect{\Psi})$ be the conditional distribution of a feature vector $\vect{x}_{i}$ given that $\vect{Z}_{i}=\vect{z}_{i}$ and $R_{i}=1$. Similarly, let $f(\vect{x}_{i} \mid R_{i}=0 ; \vect{\Psi})$  be the conditional distribution of a feature vector $\vect{x}_{i}$ given that the label $\vect{Z}_{i}$ is missing. The full semi-supervised likelihood can be written in terms of the conditional distribution given the missingness indicators,
\begin{align}
    L_{\text{full}}(\vect{\Psi};\vect{x}^{(1)}, \vect{z}^{(1)}, \vect{x}^{(2)}, \vect{r}) &= \left[ \prod_{j=1}^{n_{1}}f(\vect{x}^{(1)}_{j} \mid \vect{Z}_{j}=\vect{z}_{j},R_{j}=1 ; \vect{\Psi})\right]\left[\prod_{k=1}^{n_{2}}f(\vect{x}^{(2)}_{k} \mid  R_{k}=0; \vect{\Psi})\right] \\
    &= [L_{\text{full}}^{(1)}(\vect{\Psi}; \vect{x}^{(1)}, \vect{z}^{(1)}, \vect{r})][L_{\text{full}}^{(2)}(\vect{\Psi}; \vect{x}^{(2)}, \vect{r})].
\end{align}
If the labels are missing completely at random, $f(\vect{x}_{i} \mid \vect{Z}_{i}=\vect{z}_{i}, R_{i}=1) = f(\vect{x}_{i} \mid  \vect{Z}_{i}=\vect{z}_{i}; \vect{\Psi})$ and $f(\vect{x}_{i} \mid R_{i}=0 ; \vect{\Psi}) = f(\vect{x}_{i} ; \vect{\Psi})$. Under the conditions of Definition \ref{defn:mar}, 
    $L_{\text{full}}^{(1)}(\vect{\Psi}; \vect{x}^{(1)}, \vect{z}^{(1)}, \vect{r}) =  L_{\text{ign}}^{(1)}(\vect{\Psi}; \vect{x}^{(1)}, \vect{z}^{(1)})$ and $
     L_{\text{full}}^{(2)}(\vect{\Psi}; \vect{x}^{(2)}, \vect{r}) =  L_{\text{ign}}^{(2)}(\vect{\Psi}; \vect{x}^{(2)})$.  The objective function \eqref{eq:fsc_objective} is a weighted combination of the two correctly specified likelihood factors,
\begin{align}
    \mathcal{L}_{\text{FSC}}(\vect{\Psi} \mid  \alpha) &= [\mathcal{L}_{\text{ign}}^{(1)}(
    \vect{\Psi}; \vect{x}^{(1)}, \vect{z}^{(1)})]^{\alpha}[ \mathcal{L}_{\text{ign}}^{(2)}(\vect{\Psi} ; \vect{x}^{(2)})]^{(1-\alpha)}. \label{eq:lk_fsc}
\end{align}
If the labels are missing completely at random, assigning different weights to the likelihood blocks can be motivated using composite likelihood theory \citep{Varin2011}. The fractionally supervised classification estimator will remain consistent and asymptotically normal for all choices of $\alpha$ under mild assumptions \citep{Varin2011}. If the labels are not missing completely at random, $f(\vect{x}_{i} \mid R_{i}=1 ; \vect{\Psi}) \ne f(\vect{x}_{i} \mid R_{i}=0 ; \vect{\Psi})$, and the likelihood blocks in the fractionally supervised classification objective function \eqref{eq:lk_fsc} will no longer be correctly specified. Formally, if Definition \ref{defn:mcar} does not hold, then $L_{\text{full}}^{(1)}(\vect{\Psi}; \vect{x}^{(1)}, \vect{z}^{(1)}, \vect{r}) \ne  L_{\text{ign}}^{(1)}(\vect{\Psi}; \vect{x}^{(1)}, \vect{z}^{(1)})$ and $
     L_{\text{full}}^{(2)}(\vect{\Psi}; \vect{x}^{(2)}, \vect{r}) \ne  L_{\text{ign}}^{(2)}(\vect{\Psi}; \vect{x}^{(2)})$. In this scenario, the estimator based on maximising \eqref{eq:lk_fsc} can be highly biased for $\alpha \ne 0.5$. As discussed in \citet{Vrbik2015}, the choice of $\alpha=0.5$ is significant as the objective function can be viewed as tempered version of the ignorance likelihood \eqref{eq:ignorance}. From the Cram{\'e}r-Rao lower bound, asymptotically we expect that $\alpha=0.5$ will give the optimal fractionally supervised classification solution when the labels are missing at random. When the labels are not missing at random, the use of the pseudo-likelihood \eqref{eq:lk_fsc} requires careful consideration. 
\section{Simulation}
\label{sec:simulation}
\begin{figure}
\centering
\includegraphics[width=0.5\textwidth]{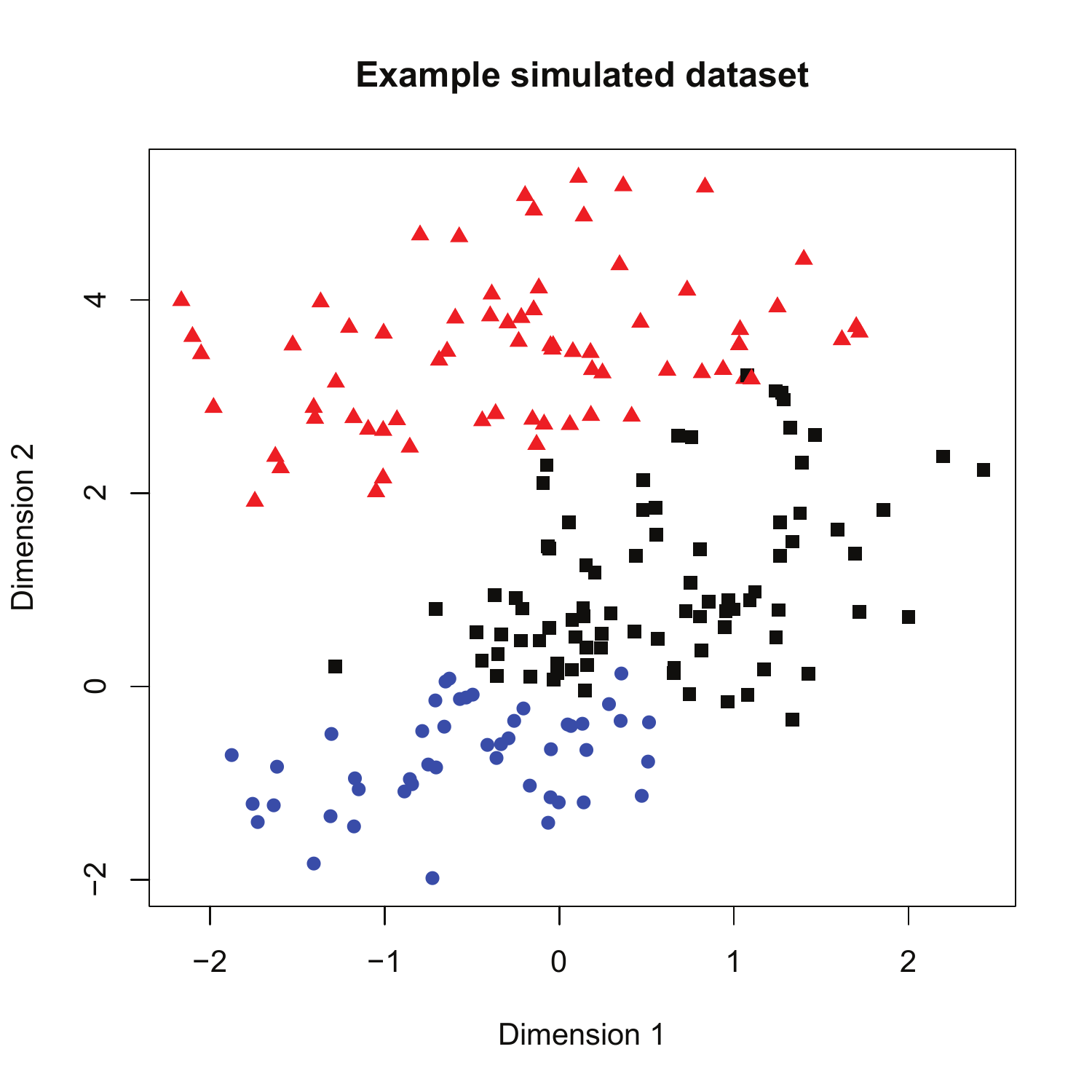}
    \caption{Example simulated dataset. Black squares denote unlabelled observations. Blue circles and red triangles are from component one and two respectively.}
    \label{fig:simulation_example}
\end{figure}

We simulated data from a two-component Gaussian mixture using the same parameter settings as in \citet{Vrbik2015}. The component means and covariance matrices were set as:
\begin{align*}
    \vect{\mu}_{1} &= \begin{bmatrix} 0 \\
    0
    \end{bmatrix}, \quad
        \vect{\mu}_{2} = \begin{bmatrix} 0 \\
    3
    \end{bmatrix}, \quad \Sigma_{1} = \begin{bmatrix}
    1 & 0.7 \\
    0.7 & 1
    \end{bmatrix},
    \quad
    \Sigma_{2} = \begin{bmatrix}
    1 & 0 \\
    0 & 1
    \end{bmatrix}.
\end{align*}
Components were equally weighted, so $\pi_{1}=\pi_{2}=0.5$. We generated $n=500$ independent observations in each simulation replication. We first generated feature vectors $\vect{x}_{1}, \ldots, \vect{x}_{n}$ and the true class membership indicators $\vect{z}_{1}, \ldots, \vect{z}_{n}$. We then set the missingness mechanism as a function of the Shannon entropy. Let $e_{i}$ denote the classification Shannon entropy of observation $i$ as per equation \eqref{eq:shannon_entropy_intro}. We introduce a random vector of missingness indicators $\vect{R}=(R_{1}, \ldots, R_{n})^{\T}$. The indicator $R_{i}$ takes value $1$ is observation $i$ is labelled, and takes value $0$ is observation $i$ is unlabelled for $i=1, \ldots n$.  We use a simple linear model for the log odds of labelling:
\begin{align}
    \log \dfrac{\Pr(R_{i}=1 \mid  \vect{x}_{i} ; \vect{\Psi}, \beta_{0}, \beta_{1})}{\Pr(R_{i}=0 \mid  \vect{x}_{i}; \vect{\Psi}, \beta_{0}, \beta_{1})} &= \beta_{0} + e_{i}\beta_{1}, \quad i=1, \ldots, n. \label{eq:simulation_log_odds}
\end{align}
If we take $\beta_{1}$ to be negative, observations that are comparatively difficult to classify are more likely to be unlabelled than observations that are comparatively easier to assign to a group. We used $\beta_{0}=1$ and ${\beta}_{1}=-5$ in the simulations. Figure \ref{fig:simulation_example} shows an example simulated dataset. Observations with missing labels are plotted as black squares. Blue circles and red triangles are from Component 1 and 2, respectively. The unlabelled observations are concentrated near the decision boundary, and the labelled observations are in regions of feature space where the classification task is very simple.

\begin{figure}
\centering
\includegraphics[width=0.8\textwidth]{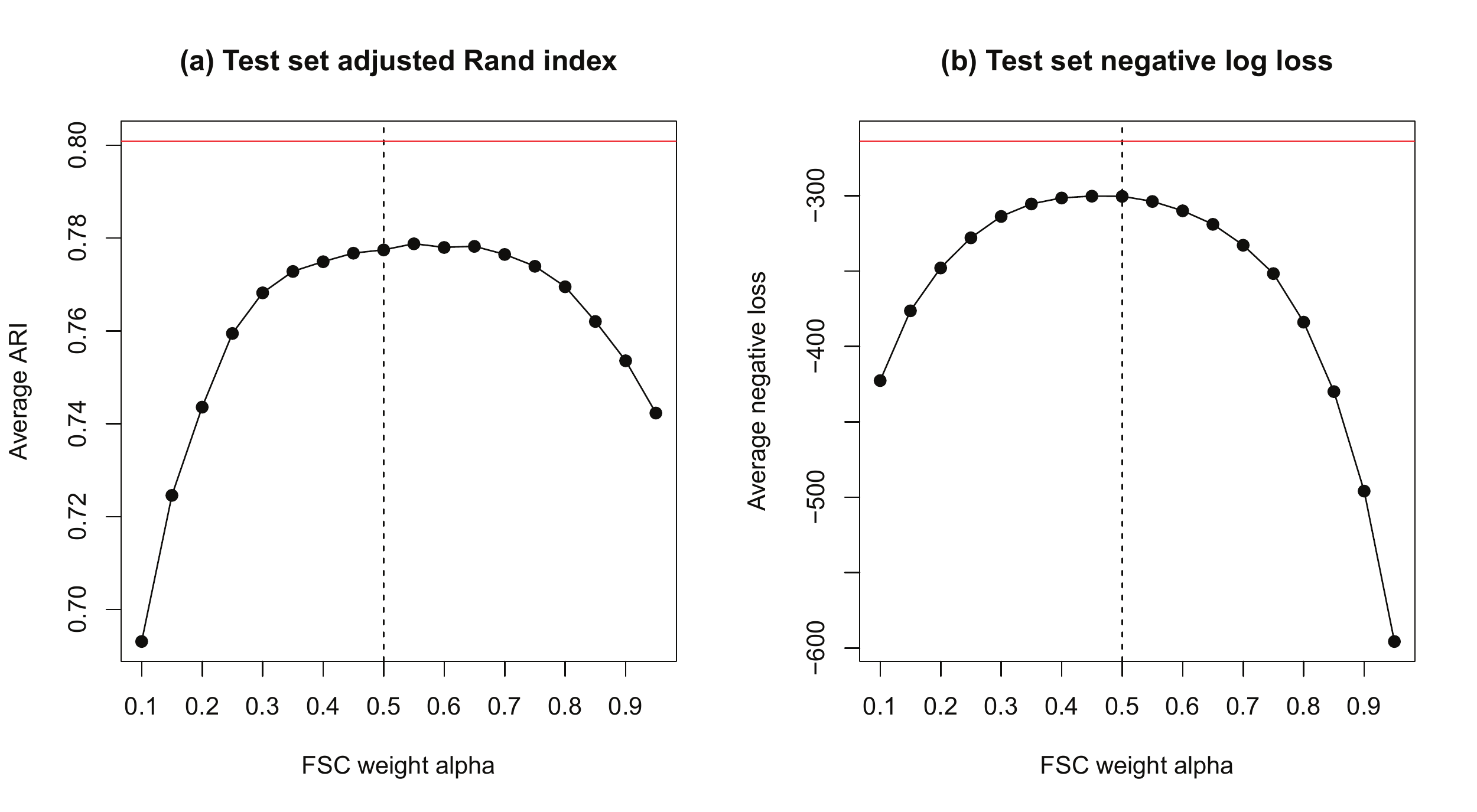}
    \caption{Simulation results. We show average results over the one hundred replications. Black points show results for fractionally supervised classification estimators at different weights $\alpha$. The red horizontal line shows results for the full likelihood estimator. Panel (a) compares the average adjusted Rand index on the independent test set. Panel (b) compares the average negative log loss on the independent test set. The joint likelihood estimator outperforms fractionally supervised classification in both performance measures.}
    \label{fig:simulation_results}
\end{figure}

We compared the different methods using the adjusted Rand index \citep{rand_1971_objective, hubert_1985}  and the negative log loss. In each simulation replication we generated a test dataset of two thousand observations.  For a particular test set, let the features and cluster indicator vector for observation $j$ be denoted as $\vect{x}_{b}'$ and $\vect{z}_{b}'$ respectively, where $b=1, \ldots, 2000$. Let $\mat{X}'$ represent the test set features and let $\mat{Z}'$ represent the test set labels. Given some parameter estimates $\widehat{\vect{\Psi}}$ we compute the predicted class memberships
\begin{align*}
    \widehat{\tau}_{bh} &= \dfrac{\widehat{\pi}_{h}f(\vect{x}_{b} ; \widehat{\vect{\theta}}_{h})}{\sum_{r=1}^{g}\widehat{\pi}_{r}f(\vect{x}_{b} ; \widehat{\vect{\theta}}_{r})}, \quad h=1, \ldots, g; \ b=1, \ldots, 2000.
\end{align*}
Let $C_{b}$ be the true cluster label for observation $b$ in the test dataset. Formally, $C_{b}=h'$ where $h'=(h \in \set{1,\ldots, g} : z_{bh}'=1)$. Let $\widehat{C}_{b}$ be the predicted cluster for observation $j$ in the test set using the fitted model:
\begin{align*}
    \widehat{C}_{b} &= \underset{h \in \set{1, \ldots, g}}{\text{argmax}} \widehat{\tau}_{bh}, \quad b=1, \ldots, 2000.
\end{align*}
We computed the adjusted rand index using the true cluster labels $C_{1}, \ldots, C_{2000}$ and the predicted cluster labels $\widehat{C}_{1}, \ldots, \widehat{C}_{2000}$. We also computed the log loss on the test set using the fitted model. Given a particular test set, let $w(\widehat{\vect{\Psi}}; \mat{X}', \mat{Z}' )$ denote the log loss using parameter estimates $\widehat{\vect{\Psi}}$:
\begin{align*}
  w(\widehat{\vect{\Psi}}; \mat{X}', \mat{Z}') &= -\sum_{j=1}^{2000}\sum_{h=1}^{g} z_{jh}' \log \widehat{\tau}_{jh}.
\end{align*}
The expected log loss is minimised using the true parameters $\vect{\Psi}$. In each simulation replication, we computed the adjusted Rand index and the log loss for each of the parameter estimates obtained using different weights $\alpha$. We also computed the adjusted Rand index and the log loss using the parameter estimate obtained by maximising the full likelihood. Figure \ref{fig:simulation_results} shows the average performance measure over the one hundred simulations. The $x$-axis is used to show the range over the different weights $\alpha$. Black points show results for fractionally supervised classification. The red line denotes the average performance of the full likelihood approach. As expected, the choice of $\alpha=0.5$ gave the best results for fractionally supervised classification as it is equivalent to use of the ignorance likelihood \eqref{eq:ignorance}. There is a notable downturn in performance when using an extreme weight close to zero or one. The full likelihood estimator had a higher average adjusted Rand index and a smaller average log loss than the fractionally supervised estimators.

\section{Conclusion}
Unlabelled observations can be an important source of information in semi-supervised learning tasks. Making full use of both unlabelled and labelled training data is an appealing idea, particular when obtaining fully labelled data is prohibitively expensive or difficult. Semi-supervised learning can be treated as a missing data problem, and analysed using the framework for missing data established by \citet{RUBIN1976}. The distinctness assumption is often seen as a mild assumption in missing data analysis \citep{schafer_1997_analysis, daniels_2015_bayesian}. However, we have found that it is an important consideration for clustering and classification. The missing label pattern in semi-supervised learning tasks can be non-ignorable due to a violation of the distinctness assumption, and the value of the unlabelled observations for statistical learning can be characterised through a logistic selection model.

Non-ignorable missing labels can arise when training set labels are given by expert assessment. Unlabelled observations are likely to be those which are difficult to confidently assign to a particular group. We found evidence of this phenomenon in a number of real datasets. This non-uniform missingness process has implications for parameter estimation. In a direct sense, model based clustering estimates class conditional densities. The estimated model also provides estimates of the decision boundaries. If unlabelled observations are known to lie near decision boundaries, the missing data pattern can be used to guide the fit of the model. We proposed a logistic selection model for the missing data mechanism that relates the difficulty of classification to Shannon entropy of the posterior class probabilities. Joint modelling involves a likelihood contribution from the pattern of missingness as well as the observed features and labels.

We used the Shannon entropy as the key base feature in the missing label model. There are many other measures that can be used to describe classification difficulty. The Shannon entropy is a special case of the R\'{e}nyi entropy, a general information theoretic measure of uncertainty. Given a a discrete probability distribution over $g$ outcomes, $p_{1}, 
\ldots, p_{g}$, the R\'{e}nyi entropy of order $\gamma$ is defined as
 \begin{align}
  H_{\gamma}(p_{1}, \ldots, p_{g}) &=  \dfrac{1}{1-\gamma} \log \left( \sum_{h=1}^{g}p_{h}^{\gamma}\right), \quad \gamma \ge 0, \gamma \ne 1.    \label{eq:renyi} 
 \end{align}
The Shannon entropy \eqref{eq:shannon_entropy_intro} is obtained by taking the limit of $H_{\gamma}(p_{1}, \ldots, p_{g})$ as $\gamma$ tends to $1$. Different values of $\gamma$ may provide more realistic models of the missingness mechanism due to selective manual labelling.

Future research directions may include an exploration of alternative missing data techniques. Observation weighting methods are a popular approach for missing data problems that do not require the construction of missing label model \citep{fitzmaurice_2015_introduction}. Fractionally supervised classification weights the log likelihood contributions of the labelled and and unlabelled observations by $\alpha$  and $(1-\alpha)$ respectively, where $0 \le \alpha \le 1$. We found that the scheme did not perform well when the labels were not missing completely at random. More sophisticated observation weighting schemes that weight observations individually may also be able to improve on the use of the ignorance likelihood \eqref{eq:ignorance}. So far we have assumed that experts are encouraged to either report an outright assignment of each observation or no classification. It is also be possible to receive a subjective probability distribution over the categories, and to model the expert labelling behaviour as a function of the model parameters \citep{aitchison_1976_diagnosis, krishnan_1990_efficiency}. It would be of interest to investigate the possible benefits of using labelling information that includes measures of expert uncertainty.

\bibliographystyle{rss}
\bibliography{bibliography}

\end{document}